\documentclass[%
 amsmath,amssymb,
 aps,
twocolumn,
pra,
floatfix,
]{revtex4-2}

\usepackage{graphicx}
\usepackage{dcolumn}
\usepackage{bm}
\usepackage{comment}
\usepackage[capitalise]{cleveref}
\usepackage[dvipsnames]{xcolor}

\newcommand{\te}[1]{\text{#1}}

\begin{document}


\title{Quantum supremacy with spin squeezed atomic ensembles}

\author{Yueheng Shi}
\thanks{These authors contributed equally}
\affiliation{State Key Laboratory of Precision Spectroscopy, School of Physical and Material Sciences,
East China Normal University, Shanghai 200062, China}
\affiliation{New York University Shanghai, 1555 Century Ave, Pudong, Shanghai 200122, China}
\affiliation{Carleton College, Northfield, MN, 55057, USA}
\affiliation{Washington University in St. Louis, St. Louis, MO, 63130, USA}

\author{Junheng Shi}
\thanks{These authors contributed equally}
\affiliation{State Key Laboratory of Precision Spectroscopy, School of Physical and Material Sciences,
East China Normal University, Shanghai 200062, China}
\affiliation{New York University Shanghai, 1555 Century Ave, Pudong, Shanghai 200122, China}

\author{Tim Byrnes}
\email{tim.byrnes@nyu.edu}
\affiliation{New York University Shanghai, 1555 Century Ave, Pudong, Shanghai 200122, China}
\affiliation{State Key Laboratory of Precision Spectroscopy, School of Physical and Material Sciences,
East China Normal University, Shanghai 200062, China}
\affiliation{NYU-ECNU Institute of Physics at NYU Shanghai, 3663 Zhongshan Road North, Shanghai 200062, China}
\affiliation{Center for Quantum and Topological Systems (CQTS), NYUAD Research Institute, New York University Abu Dhabi, UAE}
\affiliation{National Institute of Informatics, 2-1-2 Hitotsubashi, Chiyoda-ku, Tokyo 101-8430, Japan}
\affiliation{Department of Physics, New York University, New York, NY 10003, USA}



\begin{abstract}
We propose a method to achieve quantum supremacy using ensembles of qubits, using only spin squeezing, basis rotations, and Fock state measurements.  Each ensemble is assumed to be controllable only with its total spin. Using a repeated sequence of random basis rotations followed by squeezing, we show that the probability distribution of the final measurements quickly approaches a Porter-Thomas distribution. We show that the sampling probability can be related to a \#P hard problem with a complexity scaling as $ (N+1)^M$, where $ N $ is the number of qubits in an ensemble and $ M $ is the number of ensembles. The scheme can be implemented with hot or cold atomic ensembles.  
Due to the large number of atoms in typical atomic ensembles, this allows access to the quantum supremacy regime with a modest number of ensembles or gate depth. 
\end{abstract}

\maketitle


\section{Introduction} 
Quantum supremacy, or quantum computational advantage, is the notion that a quantum device can vastly outperform the computational capabilities of existing classical computers in a given task \cite{preskill2012quantum}. Obtaining a quantum speedup, as proposed in quantum algorithms such as Shor's algorithm \cite{shor1999polynomial}, has always been central to the interest in quantum computing.  Quantum simulation remains one of the most promising applications of quantum technology precisely because simulating it on a classical computer is intractable \cite{buluta2009quantum,georgescu2014quantum,cirac2012goals}. However, to demonstrate quantum supremacy, one of the important tasks is to {\it prove} the superiority of the quantum device.  The proof of this generally requires two parts \cite{Bouland2018}.  First, one requires showing that the computational task has a certain ``hardness" from the perspective of computational complexity theory, thereby invalidating the extended Church-Turing thesis. Second, one compares the computational time of the quantum device with that of the best classical algorithm running on the fastest available computer.
For the Noisy Intermediate Scale Quantum (NISQ) devices that are available today \cite{preskill2018quantum}, the limited number of qubits under imperfect conditions can be taken advantage of to design powerful classical algorithms, pushing the quantum supremacy regime to larger quantum systems.  
Current approaches for demonstrating quantum supremacy include Boson Sampling \cite{aaronson_computational_2013}, Gaussian Boson Sampling \cite{hamilton_gaussian_2017} and Instantaneous Quantum Polynomial (IQP) circuits \cite{Bremner2010,bremner_average-case_2016}.  The first experimental demonstration of quantum supremacy was achieved using random quantum circuits in a 51 qubit superconducting quantum computing device \cite{Arute2019}.  This was followed by a demonstration of quantum supremacy in Gaussian Boson Sampling \cite{zhong2020quantum,wu2021strong}, where squeezed light is input to the linear optical network. The two components, of demonstrating complexity of the problem and practical superiority to a classical algorithm, make the design of novel quantum supremacy demonstrations still quite challenging.  

Atomic systems offer a fascinating possibility in this context. They offer a high degree of controllability, and typically consist of a large number of atoms.  For example, in experiments involving Bose-Einstein condensates one typically has $ \sim 10^5 $ atoms \cite{greiner2002quantum}, and for experiments with atomic ensembles typically have $ \sim 10^{12} $ atoms \cite{julsgaard2001experimental}. While this far exceeds the number of qubits in state-of-the-art quantum computers, one of the limitations is the lack of microscopic control of the atoms, although in recent years progress on front has also been made \cite{bernien2017probing}. For this reason one of the main applications of such systems has been in the context of quantum simulation \cite{buluta2009quantum,georgescu2014quantum,cirac2012goals}, where microscopic control is often unnecessary to realize a physical model. 
In Ref. \cite{kocharovsky2022quantum} a proposal was made based on sampling of the Bogoliubov distribution in a multi-trap cold atom system. Up to this point it has remained a tantalizing possibility to rigorously show that atomic systems also lie in the quantum supremacy regime.

\begin{figure}[t]
\includegraphics[width=8cm]{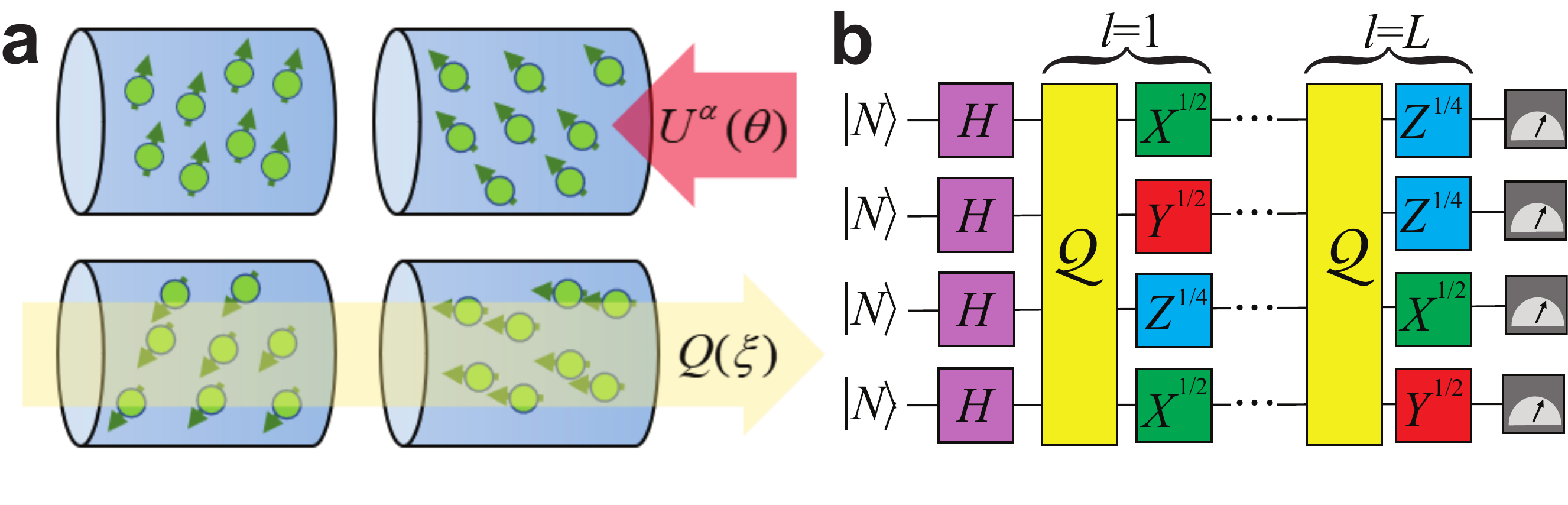}
\caption{Random quantum circuits using qubit ensembles. (a) The schematic setup considered in this paper.  $ M $ ensembles of qubits, each containing $ N $ qubits are controlled using basis rotations $ U^\alpha (\theta) $ as defined in (\ref{basisrot}) and squeezing operations $ Q_{nm} (\xi) $.  In this example $ N = 8 $ and $ M = 4$.  (b) An example of a random quantum circuit for $ M = 4$.  Each horizontal line denotes a qubit ensemble with $ N $ qubits.  $ L $ cycles, consisting of a randomly chosen basis rotation followed by a squeezing operation, followed by a measurement in the Fock basis constitutes the random circuit. }
\label{fig1}
\end{figure}

In this paper, we propose an experimental scheme to achieve quantum supremacy with ensembles of qubits, using only spin squeezing, basis rotations, and total spin measurements.  The basic scheme is shown in Fig. \ref{fig1}. After initializing the qubits in a spin coherent state \cite{Byrnes2021}, they are spin squeezed in random bases, by applying a sequence of spin squeezing and rotations around the $ x,y,z $-axes.  The aim is then to perform sampling of the measurement distribution for a given squeezing sequence.  We analyze the complexity of simulating such random circuits classically, and show that this is intractable for large particle and ensemble numbers, by connecting it to a \#P-hard problem. Finally, we show a classical simulation method suitable for large scale systems and show the regime in which quantum supremacy should be attainable.

\section{Physical system}

We consider $ M $ ensembles each containing $ N $ qubits (Fig. \ref{fig1}(a)). Such qubit ensembles can be implemented using hot atomic ensembles in glass cells, or optically/magnetically trapped cold atoms.  Suitable logical states are the hyperfine ground states of the atoms \cite{hammerer2010,pezze2018}.  For atomic ensemble in glass cells, the number of atoms can be typically in the region of $ N \sim 10^{12} $ \cite{pezze2018,hammerer2010, julsgaard2001experimental,Bao2020}. Multi-ensemble systems have been realized with atomic cells, in Ref. \cite{pu2017experimental,pu2018experimental} $ M = 25, 225 $ was achieved.  We assume that the qubits within each ensemble cannot be individual controlled, but each ensemble can be addressed individually.  The only operators that are available for control and readout of the ensembles are in terms of collective spin operators, defined as $ \hat{S}^\alpha_m = \sum_{n=1}^N \hat{\sigma}_{n,m}^{\alpha} $, where $ \alpha \in \{x,y,z \}$,  $ m \in [1,M]$ labels the ensemble, and $ \hat{\sigma}_n^{\alpha}$ are the Pauli matrices.  Spin operators satisfy commutation relations $ [\hat{S}^\alpha_m, \hat{S}^\beta_{m'} ] = 2i \epsilon^{\alpha \beta \gamma} \delta_{m m'} \hat{S}^\gamma_m  $, where $ \epsilon^{\alpha \beta \gamma}$ is the Levi-Civita antisymmetric tensor and $ \delta_{mm'} $ is the Kronecker delta.  

Our quantum supremacy protocol is based on a combination of quantum gates on the ensembles, followed by a measurement. 
The following operations are assumed to be available to realize quantum supremacy in a such a system of ensembles.  First, we assume that we can perform basis rotations $  U^\alpha_n (\theta) = e^{-i \hat{S}^\alpha_m \theta} $ on each ensemble. 
These are collective spin rotations, and correspond to the  simultaneous rotations of all $N$ individual qubits about the same axis. Such collective spin rotations are routinely performed with either radio frequency/microwave or Raman pulses in atomic ensembles \cite{Byrnes2021,Abdelrahman2014,genov2014correction}.  
In fact for our random quantum circuit we make the further restriction to the following gates
\begin{align}
{\hat{X}}_m^{1/2}  = e^{-i\hat{S}_m^x\pi/4} \hspace{5mm}
{ \hat{Y} }_m^{1/2} = e^{-i\hat{S}_m^y\pi/4} \hspace{5mm}
{ \hat{Z} }_m^{1/4} = e^{-i\hat{S}_m^z\pi/8} .  
\label{basisrot}
\end{align}
Note that the $ z $-axis rotation is taken to correspond to the $ \pi/8$ gate, which is to ensure that a non-Clifford gate is present for basis rotations \cite{gottesman1998heisenberg}.  While this is not essential since the squeezing gates are non-Clifford gates, this helps to improve the convergence of the random circuit.  

We also assume that spin squeezing operations are available on the ensembles. A squeezing operation applied on the $ m$th ensemble is defined using the operator $ \hat{Q}_m (\xi) = e^{-i(\hat{S}_m^z)^2 \xi } $ which is a one-axis spin squeezing Hamiltonian \cite{kitagawa1993squeezed}.
The squeezing operation produces correlations between the particles, and is an entangling operation \cite{sorensen2001many}.  
Such squeezing has been realized in cold atoms using nonlinear interactions \cite{gross2012spin,riedel2010atom} or quantum nondemolition measurements in hot atomic ensembles \cite{hald1999spin,kuzmich2000generation,hammerer2010,takano2009spin,Bao2020}.  
By considering multiple ensembles as a single spin, spin squeezing across multiple ensembles may be produced $  \hat{Q}_{nm} (\xi) = e^{-i(\hat{S}_m^z+ \hat{S}_n^z)^2 \xi }$.  
This squeezing operation generates entanglement between different ensembles due to the cross terms $ \hat{S}_m^z \hat{S}_n^z $ \cite{byrnes2013fractality,jing2019split,kitzinger2020two}. Considering all the ensembles together produces squeezing across all the ensembles according to
\begin{equation}
\hat{\cal Q}(\xi) = e^{-i (\sum_{m=1}^M \hat{S}^z_m)^2 \xi } . 
\label{squeezingall}
\end{equation}
Squeezing across multiple ensembles has been demonstrated in hot atomic ensembles using quantum nondemolition measurements \cite{julsgaard2001experimental,hammerer2010,krauter2013deterministic}.  Several proposals for producing squeezing between cold atom ensembles have been proposed \cite{pyrkov2013entanglement,PhysRevA.74.022312,aristizabal2021quantum,pyrkov2014full,jing2019split,byrnes2013fractality,rosseau2014entanglement}.  

Finally, the measurement is performed in the eigenbasis of the $ \hat{S}^z_m $ operator, defined as  
\begin{align}
\hat{S}^z_m | k \rangle = (2k - N) | k \rangle , \label{fockstates}
\end{align}
where $ k \in [0,N] $.

\section{Random quantum circuits}

Our approach to quantum supremacy follows a similar general approach to  random quantum circuits as demonstrated in Ref. \cite{Arute2019}.  We first initialize all the ensembles to a spin coherent state polarized in the $ z $-direction $  |\psi_0 \rangle = | k = N \rangle^{\otimes M }  = | 0 \rangle^{\otimes NM} $.  The first step of the circuit is to apply a Hadamard gate on each ensemble $ H_m = U^x_m (\pi/2) U^z_m (\pi/2) U^x_m (\pi/2)$. This produces an equal superposition of all  states $ |+ \rangle^{\otimes NM} $.  
The random circuits we consider consist of $L$ cycles, where each cycle contains two gates consisting of the squeezing operation (\ref{squeezingall}) followed by a basis rotation (\ref{basisrot}).  After $L$ such cycles, a projection measurement is made on the Fock basis (\ref{fockstates}), obtaining a single sample 
(Fig. \ref{fig1}(b)).  The aim of the quantum circuit is to obtain the probability distribution of a given measurement outcome, given by 
\begin{align}
p_{\vec{k} } & = | \langle \vec{k} | {\cal C } |\psi_0 \rangle |^2 ,  \label{probabilityk} 
\end{align}
where the random quantum circuit is
\begin{align}
{\cal C } & = \left( \prod_{l=1}^L   \hat{U}_l \hat{{\cal Q}} \right)  H^{\otimes M } .  
\label{randomqcircuit}
\end{align}
Here $ \hat{U}_l = \otimes_{m=1}^M \hat{W}_m $ and $ \hat{W}_m $ is randomly chosen from $\{ \hat{X}_m^{1/2}, \hat{Y}_m^{1/2}, \hat{Z}_m^{1/4} \} $ on each cycle $ l  $.  A particular measurement outcome is given by $ | \vec{k} \rangle = \otimes_{m=1}^M | k_m \rangle $, 
where $\vec{k} = (k_1, \dots, k_M)  $.

\section{Randomness of quantum circuit}

\begin{figure}[t]
\includegraphics[width=\columnwidth]{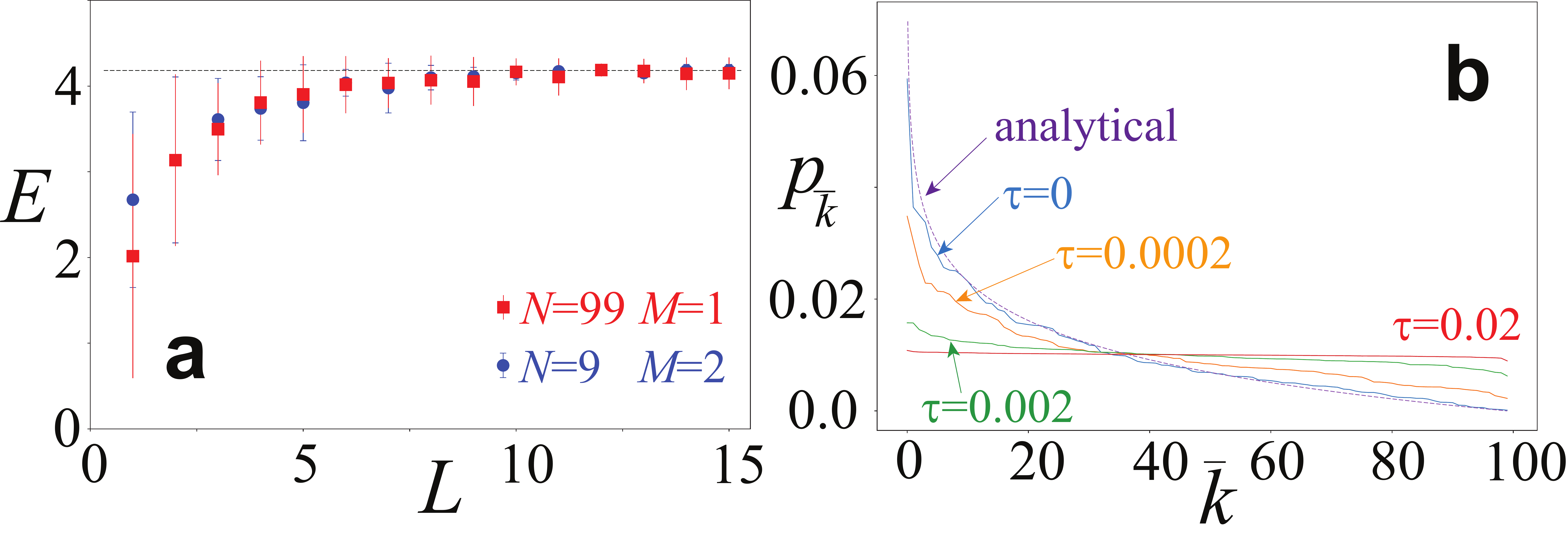}
\caption{Randomness of the measured probability distribution for the random quantum circuit (\ref{randomqcircuit}).  (a) Entropy $ E = - \sum_{\vec{k}} p_{\vec{k}} \ln p_{\vec{k}} $ of the probability distribution (\ref{probabilityk}).  We show two cases with (i)  $ M = 1, N = 99$; (ii) $ M =2, N = 9$. (b) Sorted probability distribution (\ref{probabilityk}) after $ L =10$ cycles for a $ M =1$ and $ N = 99$ system.  $\bar{k} $ is the sorted label.  Solid lines are for different rates of Lindbladian dephasing as marked, applied after each cycle. Dashed line is the ideal PT-distribution  $ p_{\bar{k} } =(-\ln(\bar{k})+\ln[(N+1)^M])/(N+1)^M$. In all cases a squeezing parameter of $ \xi = \pi/ \sqrt{MN} $ is chosen.  
\label{fig2}}
\end{figure}

To verify that our proposed sequence gives a Gaussian random state, we calculate the entropy of the measurement probabilities (Fig. \ref{fig2}(a)). For a $D\times D$ random matrix, the  probability follows the Porter-Thomas (PT) distribution, which has an entropy of $ \ln D-1+\gamma$, where $\gamma \approx 0.577$ is the Euler constant\cite{porter_fluctuations_1956}. Fig. \ref{fig2}(a) shows the entropy of the probabilities (\ref{probabilityk}) as a function of number of cycles. Even for the relatively small number of particles, we observe a fast convergence to the predicted entropy of the PT-distribution after typically $ L = 8 $ cycles. We also verify that the probability distribution (\ref{probabilityk}) follows the PT-distribution, by showing the sorted probabilities after $L = 10 $ cycles (Fig. \ref{fig2}(b)).  We see that for the probabilities (\ref{probabilityk}) there is excellent agreement, again despite the small number of particles examined. We also examine the effect of decoherence on our proposed sequence, by applying a Lindbladian dephasing after each cycle (see Appendix).  The general effect is to turn the PT-distribution into a uniform distribution, which is as expected as the state becomes a completely mixed state.

\section{Complexity analysis} 

We now provide a proof of the hardness of our problem in terms of computational complexity theory. As given in previous works \cite{aaronson_computational_2013,bremner_classical_2011,bremner_average-case_2016,bremner_classical_2011}, the complete proof to show that there exists no efficient classical sampler to simulate the output of an average-case circuit follows several steps.  The first step is to show the existence of a class of worst-case circuits and prove that the complexity of estimating the probability of certain outputs in those circuits is \#P-hard. The second step is to extend this hardness to average-case circuits by a worst-to-average reduction. Bouland, Vazirani and co-workers showed a general method on how to perform a reduction from the worst-case circuit to the average-case \cite{Bouland2018}. 
The third step is to show if an efficient classical algorithm that can approximate the output probability of an average-case circuit up to an additive error, then the polynomial hierarchy will collapse to its third level \cite{arora_computational_2009,toda_pp_1991,aaronson_quantum_2005,han_threshold_1997}.  As the second and third steps are established in previous papers, we focus our attention on the first step (see Appendix for a summary).  

\begin{figure}[t]
\includegraphics[width=8cm]{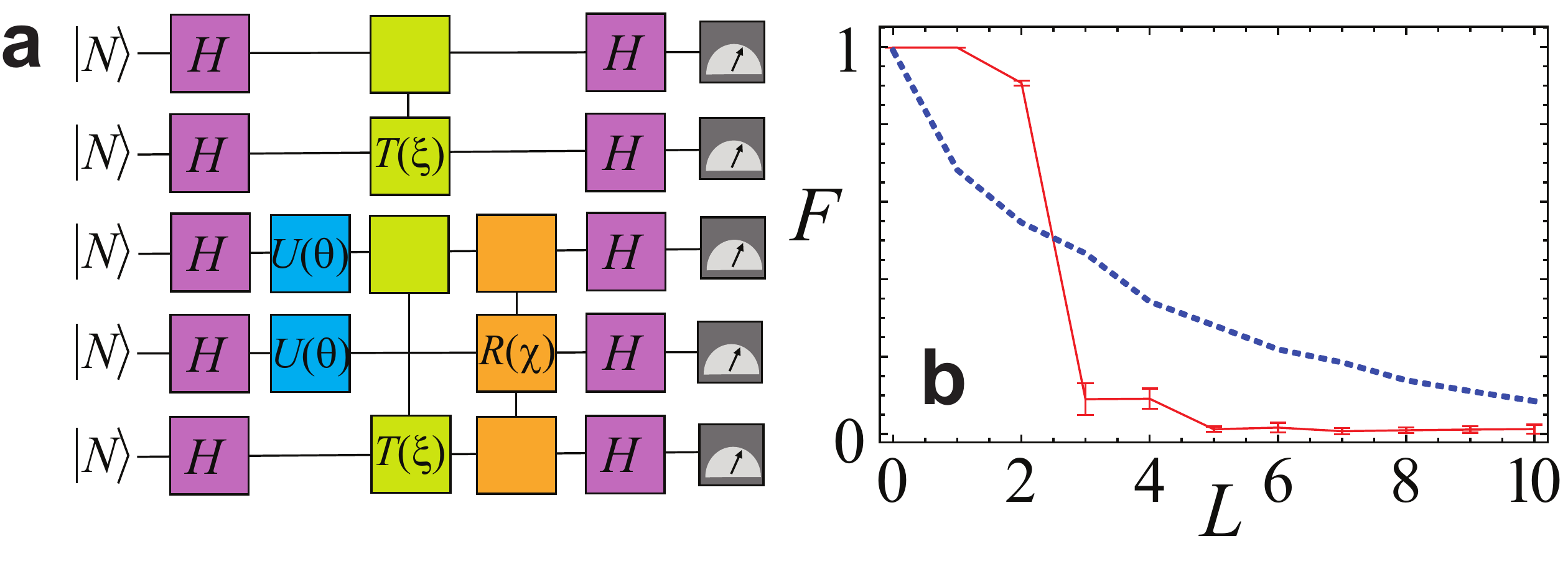}
\caption{(a) Example of the worst-case circuit used in the evaluation of the probability (\ref{probzero}).  The $ T , R $ gates are the two and three ensemble interactions (\ref{squeezingthree}) respectively, and the boxes indicate the ensembles that interact.  (b) Path integral Monte Carlo evaluation of the fidelity of the final state (solid lines), using (\ref{pathintens}). Calculation uses $ N = 99 $, $ M = 1 $ with a gate sequence $ \{ X^{1/2}, Y^{1/2}, Z^{1/4}, X^{1/2}, Z^{1/4},Y^{1/2},Z^{1/4},X^{1/2}, Y^{1/2},X^{1/2} \} $ with $10^5$ sampled paths.  For comparison the fidelity using a direct matrix evaluation (dashed lines) with dephasing time $ \tau = 10^{-4} $ after each cycle is shown.  }
\label{fig3}
\end{figure}

Consider the class of quantum circuits given in Fig. \ref{fig3}(a).  After the initial Hadamard gates,
commuting two- and three-ensemble interactions of the form
\begin{equation}
\hat{T}_{nm}(\xi) = e^{-i\hat{S}_n^z \hat{S}_m^z  \xi }, \hspace{1cm} 
    \hat{R}_{lmn} (\chi) = e^{-i\hat{S}_l^z \hat{S}_m^z \hat{S}_n^z \chi }  
    \label{squeezingthree}
\end{equation}
are performed, as well as basis rotations around the $ z$-axis.  Such gates can be produced by universality arguments \cite{lloyd1995almost,byrnes2015macroscopic} (see Appendix).  Finally a Hadamard gate is applied, followed by a measurement.  The probability of the outcome $ \vec{k} = \vec{0} $ can be calculated using standard methods (see Appendix) to be
\begin{align}
 p_{\vec{0}}  = & \frac{\left|  \sum_{\vec{\sigma}} 
e^{-i \pi f(\vec{\sigma})} \right|^2 }{4^{NM}} =\frac{\left|  \sum_{\vec{\sigma}} 
(-1)^{f(\vec{\sigma})} \right|^2 }{4^{NM}}  \label{probzero} \\
 f(\vec{\sigma}) = & \sum_{m_1  m_2  m_3}  \alpha_{m_1 m_2 m_3} k_{m_1} k_{m_2} k_{m_3} \nonumber \\
& + \sum_{m_1 m_2} \beta_{m_1 m_2}  k_{m_1} k_{m_2} + \sum_m  \gamma_m  k_{m} .\label{hardness_f}
\end{align}
where $ k_m = (N + \sum_{n=1}^N \sigma_{n,m})/2 $ counts the number of $  \sigma_{n,m} =1$ in the $ m$th ensemble. The sum over $ \vec{\sigma} $ runs over the  $ 2^{NM} $ configurations of the whole system.  The parameters $  \alpha, \beta, \gamma$ can be simply related to the evolution times of the circuit in Fig. \ref{fig3}(a), such that any desired set of parameters can be created. Thus using a suitable circuit it is possible to realize $ \alpha_{m_1 m_2 m_3}, \beta_{m_1 m_2}, \gamma_m \in \{0, 1\} $.  The sum in (\ref{probzero}) is known as the gap function, and when $ f $ is a degree 3 polynomial, it is known to be \#P-hard to calculate for the case that $ k_m \in \{0,1\} $ \cite{bremner_average-case_2016,Gao2017}. 
The difficulty of the evaluation of the sum originates from the lack of simple structure of $ (-1)^{f(\vec{k})} $, such that the number of $ \vec{k}$ that give $ \pm 1 $ cannot be found easily. In our case, $ k_m \in [0,N] $, rather than $ k_m \in \{0,1\} $, but due to the fact that only the parity of the function $ f $ matters in the sum in (\ref{probzero}), it follows that only the parity of the $ k_m $ matters for the function $ f $. For example, for a term such as $ (-1)^{k_1 k_2 k_3} $, this is only $-1$ when $ k_1, k_2, k_3 $ are all odd.  
This means that here $ f $ encodes the same problem as the binary case, with the mapping $ k_m \rightarrow k_m \text{mod } 2$.  This shows that sampling from the circuit in Fig. \ref{fig3}(a) is \#P-hard, by equivalence to the original binary case.  

Another way to see the complexity of circuits of the form of Fig. \ref{fig3}(a) is by connecting it to IQP.  Using an extension of the arguments used to derive (\ref{squeezingthree}), one may show that any gate of the form $\exp\left[-i\theta \prod_{m\in \mathcal{M}}\hat{S}_m^z \right]$ can be generated, where $\mathcal{M}$ runs over a subset of the ensembles.  For the $ N = 1 $ case, using such gates in a circuit of the form of Fig. \ref{fig3}(a) coincides exactly with IQP. We may then use the results of Ref. \cite{fujii_commuting_2017},  which showed the hardness of IQP by connecting it to the hardness of calculating the partition function. For $ N>1 $, the extremal values $ k = \{0,N \} $  coincide with the $ N = 1$ case \cite{mohseni2021error}, but the sum in the probability expression will involve additional terms that are distinct to the extremal values. Thus for an exact evaluation of the amplitude, the complexity of the circuit is at least as hard as IQP.


\section{Simulation algorithm}

Demonstrating quantum supremacy requires not only realizing a quantum device that performs sampling efficiently, but also comparison to a classical algorithm that can perform the corresponding calculation.  Here we describe an algorithm using a Feynman Path Integral (FPI)  classical sampling approach \cite{Boixo2017,Boixo2018}. 
The aim of the FPI based sampling algorithm is to calculate the amplitude 
\begin{align}
\langle \vec{k}_f | \hat{{\cal C}} | \psi_0 \rangle = 
\sum_{ \vec{k}^1, \dots, \vec{k}^{T-1} } \prod_{t=1}^T 
\langle \vec{k}^{t} | C^{(t)}| \vec{k}^{t-1} \rangle , 
\label{pathintens}
\end{align}
where $ | \vec{k}^t \rangle= \otimes_{m=1}^M | k_m^t \rangle $ is a ensemble configuration with $ k_m^t \in [0,N] $ and $ {\hat{\cal C}} = \prod_{t=0}^T  C^{(t)} $ is the total circuit written gate by gate.  The number of the gates applied in total is $T$, the initial state is $ | \vec{k}^0 \rangle = | \psi_0 \rangle $, and the final state is $ | \vec{k}_f \rangle = | \vec{k}^T \rangle $. Each term in the multidimensional sum of (\ref{pathintens}) represents a path in spin configuration space, evolving in a ``time'' direction labeled by $ t $.  The sum in (\ref{pathintens}) in principle runs over $ (N+1)^{M(T-1)} $, terms but due to the diagonal nature of the $ {\cal Q} , \hat{Z}_m^{1/4} $ gates, and the local nature of the $ \hat{X}_m^{1/4}, \hat{Y}_m^{1/4}$,  many of the amplitudes give zero (see Appendix).  This reduces the sum to $ (N+1)^G $, where $ G $ is the number of two-sparse gates (i.e. the off-diagonal $ \hat{X}_m^{1/4}, \hat{Y}_m^{1/4}$). 

For large $ N $, this can still be a formidable number of paths, and hence one can approximate the amplitude by randomly sampling from these paths.  Such a path integral Monte Carlo approach is effective when the Hilbert space dimension is prohibitively large to perform a direct matrix computation, which scales in complexity as $ G (N+1)^{2M}$.  For ensemble sizes $ N = 10^{12} $ and $ M = 2$ such an exact computation is not tractable.  On the other hand, it is possible always to perform a Monte Carlo calculation, by simply adjusting the number of paths, at the cost of a reduced fidelity.  Fig. \ref{fig3}(b) shows an example of the fidelity of the FPI method with the number of cycles for a fixed number of paths. 
We compare this to the fidelity of the direct matrix calculation including decoherence, which also shows a decrease in fidelity with $L $, since the dephasing is added after each cycle.  We see that for small $ L $, the FPI method produces good estimates of the state, but sharply loses fidelity when $(N+1)^G $ exceeds the number of sampled paths. Meanwhile, the decoherence calculation decreases but at a slower pace, such that some fidelity is retained even for deep circuits. We expect a similar situation for an experimental quantum supremacy demonstration, where the loss of fidelity for the experiment will have a slower decay with $ L $ than the FPI approach. In Ref. \cite{Arute2019}, it is estimated that a million CPU cores could be used to sum over $\sim 10^{14}$ paths over 2 weeks. Hence for $ N =10^{12}$ and $ G = 2$ we expect that the fidelity of the FPI will be very low, giving opportunity for a quantum device to exceed its performance.

\section{Conclusions}

We have proposed a route to achieve quantum supremacy using ensembles of qubits, where each ensemble is controlled only using its total spin. The primary physical platform for realization of this scheme is atomic ensembles, where $ N $ can be extremely large.  The relevant Hilbert space for the system is $(N+1)^M$, meaning that even for a modest number of ensembles, the complexity can be enormous.  Despite the lack of microscopic control of the atoms, the system quickly approaches a PT-distribution.  We showed the complexity of the problem by showing that a hard instance of the circuit is equivalent to a \#P-hard problem.  A path integral simulation algorithm was also introduced, which is appropriate for simulating large-scale systems, and has a complexity scaling as $ (N+1)^G$. 

Despite only having limited control of the quantum system, i.e. no microscopic control of individual qubits, it is possible to connect it to a computationally difficult problem.  Having limited control is one of the principal technological differences in a quantum simulator versus a quantum computer, but we see here that nevertheless that the complexity of classically simulating such devices can be high. Using a similar ensemble approach, quantum computing has been proposed \cite{byrnes2012macroscopic,mohseni2021error,abdelrahman2014coherent,Byrnes2021}, suggesting further applications beyond quantum supremacy. We anticipate the  primary technological difficulty in realizing our scheme is in the measurement readout, where ideally the spin of the ensemble should be read out with single atom resolution.  In the context of cold atoms, advances have been made where close to single atom resolution has been achieved \cite{hume2013accurate,huper2019preparation}. Analogous challenges have been present in the optical context, where the lack of single photon resolution has shown not to be an impediment towards reaching quantum supremacy \cite{zhong2020quantum,wu2021strong,quesada_gaussian_2018,shi2021gaussian}.


\begin{acknowledgments}
This work is supported by the National Natural Science Foundation of China (62071301); NYU-ECNU Institute of Physics at NYU Shanghai; the Joint Physics Research Institute Challenge Grant; the Science and Technology Commission of Shanghai Municipality (19XD1423000,22ZR1444600); the NYU Shanghai Boost Fund; the China Foreign Experts Program (G2021013002L); the NYU Shanghai Major-Grants Seed Fund.
\end{acknowledgments}


\appendix
\section{Completely symmetric subspace}

In this section, we show the mapping between the completely symmetric qubit states and bosonic Fock states on a single ensemble.  

The Hilbert space on each ensemble is formally of dimension $ 2^N$.  However, the collective spin operators are symmetric under particle interchange, and we will also start in an initial state that obeys this symmetry.  Under these conditions, it is possible to restrict the Hilbert space to a smaller subspace, where all states are symmetric under interchange \cite{Byrnes2021}.  In this case we may map the total spin operators to Schwinger boson operators
\begin{align}
\hat{S}^x_m & = \hat{a}_m^\dagger\hat{b}_m+\hat{b}_m^\dagger\hat{a}_m \nonumber \\
\hat{S}^y_m & = -i\hat{a}_m^\dagger\hat{b}_m+i\hat{b}_m^\dagger\hat{a}_m \nonumber \\
\hat{S}^z_m & = \hat{a}_m^\dagger\hat{a}_m-\hat{b}_m^\dagger\hat{b}_m, 
\end{align}
which act on orthonormal Fock states are defined as 
\begin{align}
|k\rangle = \frac{(a^\dagger_m)^k(b^\dagger_m)^{N-k}}{\sqrt{k!(N-k)!}} | \text{vac} \rangle ,
\end{align}
where $\hat{a} $ and $\hat{b} $ are bosonic annihilation operators. Thus for a single qubit ensemble, there are $N+1$ Fock states available, and the total Hilbert space of the $ M $ ensemble system has a dimension of $ (N+1)^M $.

\section{Ensemble gates}

In this section we show how to produce two- and three-ensemble entangling gates. It also serves as an example for demonstrating the construction of unitary commuting gates which are the key component of IQP circuits in our system. Such commuting gates can be produced by universality arguments as we show below.  

\subsection{Two-ensemble interactions}

Expanding two-ensemble squeezing gate, we see that it is a combination of an entangling interaction as well as squeezing on individual ensembles
\begin{align}
\hat{Q}_{nm} (\xi) & = e^{-i ((\hat{S}^z_n)^2 + (\hat{S}^z_m)^2 + 2 \hat{S}^z_n \hat{S}^z_m) \xi} . 
\end{align}
By applying local squeezing operators we may obtain the two-ensemble interaction gate
\begin{align}
T_{nm} (\xi) & = \hat{Q}_{nm} (\xi/2) \hat{Q}_{n} (-\xi/2) \hat{Q}_{m} (-\xi/2) \nonumber \\
& = e^{-i  \hat{S}^z_n \hat{S}^z_m \xi} . 
\end{align}

\subsection{Three-ensemble interactions}

In this section we show that using the assumed operations of the main text, it is possible to produce effective three-ensemble interactions.  

First consider the evolution sequence
\begin{align}
    U^A_{nm} & = e^{i \hat{S}^y_n \pi/4} \hat{Q}_{nm} (\phi)  \hat{Q}_n (- \phi) \hat{Q}_m (-\phi)  e^{-i \hat{S}^y_n \pi/4}  \nonumber \\
    & = e^{i \hat{S}^y_n \pi/4} e^{i ( \hat{S}^z_n + \hat{S}^z_{m})^2  \phi } 
    e^{ -i ( ( \hat{S}^z_n )^2 +( \hat{S}^z_m )^2) \phi } 
    e^{-i \hat{S}^y_n \pi/4} \nonumber \\
     & = e^{i \hat{S}^y_n \pi/4} e^{2 i \hat{S}^z_n \hat{S}^z_{m} \phi } 
    e^{-i \hat{S}^y_n \pi/4} \nonumber \\   
&=  e^{2 i
 \hat{S}^x_n \hat{S}^z_m \phi }  . 
\end{align}
We can see that the above sequence produces the effective Hamiltonian 
\begin{align}
H_{nm}^A = \hat{S}^x_n \hat{S}^z_m .  
\label{hama}
\end{align}
Similarly defining the same sequence but with initial rotations around the $ x $-axis we have 
\begin{align}
    U^B_{nm} & = e^{i \hat{S}^x_n \pi/4}\hat{Q}_{nm} (\phi)  \hat{Q}_n (-\phi) \hat{Q}_m (-\phi)   e^{-i \hat{S}^x_n \pi/4}  \nonumber \\
&=  e^{2 i
 \hat{S}^y_n \hat{S}^z_m \phi }  . 
\end{align}
which gives rise to the Hamiltonian
\begin{align}
H_{nm}^B = \hat{S}^y_n \hat{S}^z_m .
\label{hamb}
\end{align}

We then use the general result of Ref. \cite{lloyd1995almost}, where it is shown that 
\begin{align}
e^{[A,B] t} \approx \left( e^{-iB \sqrt{t/n} } e^{-iA  \sqrt{t/n} }  e^{iB \sqrt{t/n} } e^{iA  \sqrt{t/n} }  \right)^n ,
\label{commutatorlloyd}
\end{align}
where $ A, B $ are Hamiltonians that are available and the approximation improves for large $ n $.  The meaning of this is that if Hamiltonians $ A, B $ are available, then it is also possible to implement the Hamiltonian $ i [A,B] $.  

Using the commutator relation (\ref{commutatorlloyd}) we can see that using (\ref{hama}) and (\ref{hamb}) it is possible to produce the three-ensemble interaction
\begin{align}
H_{nml}^{(3)} & = i [H_{nm}^A, H_{nl}^B] \nonumber \\
& = - 2  \hat{S}^z_n  \hat{S}^z_m  \hat{S}^z_{l}  ,
\end{align}
as desired.

\subsection{Higher order interactions}

Higher order interactions can be produced by the same arguments as for three-ensemble interactions.  For example, by commuting Hamiltonians (\ref{hama}) and 
\begin{align}
    H_{nml}^B =  \hat{S}^y_n  \hat{S}^z_m  \hat{S}^z_{l} 
\end{align}
we may generate fourth order Hamiltonians, and the process can be repeated.

\section{Dephasing evolution}

We apply a dephasing evolution of Lindblad form to examine the effect of decoherence on the Porter-Thomas distribution.  The master equation reads
\begin{equation}
    \frac{d\rho}{d t} = -\frac{i}{\hbar}[H,\rho]+
    \gamma\sum_{m=1}^M (\hat{S^z_m}\rho \hat{S^z_m}-\frac{1}{2}\{ (\hat{S^z_m})^2,\rho\}) . 
\end{equation}
We apply the master equation after each cycle, i.e. after application of $ \hat{U}_l \hat{\cal Q} $.  In the case that $ H = 0 $, the master equation can be solved exactly, such that after evolving for a time $ t $, the density matrix becomes
\begin{align}
\rho_{\vec{k} \vec{k}'} (\tau) = e^{-2 \tau \sum_{m=1}^M (k_m- k_m')^2 } \rho_{\vec{k} \vec{k}'} (0)
\end{align}
where $ \tau = \gamma t $ and the density matrix elements are defined as
\begin{align}
 \rho_{\vec{k} \vec{k}'}  (t) & = \langle \vec{k} | \rho(\tau ) | \vec{k}' \rangle 
\end{align}
and $\vec{k} = (k_1, \dots, k_M)  $.

\section{Computational complexity proof}

In this section, we provide the complexity proof sketched in the main text. First, we elaborate the demonstration of the worst-case circuit in our scheme such that exactly computing its probability amplitude is $\#$P-hard. Next, we provide a detailed sketch of proof for the second step and third step mentioned the main text. The second step is for the worst-to-average reduction proof employed in Ref. \cite{Bouland2018} to show that approximating the output probability of an average-case circuit in our scheme is $\#$P-hard too. The third step is relating an individual output probability to simulating the whole distribution mainly with the help from Stockmeyer Counting Theorem. Then we disapprove the existence of an efficient classical sampler that can approximate the distribution to an additive error, otherwise the polynomial hierarchy will collapse to its third order.

\subsection{Worst-case circuit}
\label{sec:worstcaseconj}

We consider a similar approach of constructing the worst-case circuit as in Refs. \cite{bremner_average-case_2016}. In their approach, the output probability of the constructed circuit is mapped to a function which is known to be computationally hard. More specifically, they find a circuit $\hat{\cal C}_\te{H}$ such that 
\begin{equation}
     \langle \vec{0} |\hat{\cal C}_\te{H} | \psi_0 \rangle  \propto \sum_z (-1)^{f(z)} , 
    \label{previousres}
\end{equation}
where $z$ is a binary string of length $ M $ and $f(z)$ is a degree 3 polynomial that maps $z$ to integers, namely $f(z) \in \mathcal{Z}$. The right hand side of (\ref{previousres}) is known to be a $\#$P-hard function to compute exactly since it corresponds to the counting problem of the number of $z$ whose $f(z)$ is even minus the number of $z$ whose $f(z)$ is odd. This might seem different from the original definition where $f(z)\in \{0,1\}$. But as pointed out in the main paper, only the parity of $f(z)$ matters, such  that the mapping $z \rightarrow z \mod 2$ makes the problems equivalent. While the initial state is fixed, the final state is dependent upon the measurement outcome. Due to the ``hiding" property of RCS circuit \cite{Bouland2018}, we can however focus on a fixed output. In this way, it is argued that the random circuit $\hat{\cal C}_\te{H} $ is a problem \#P-hard. Intuitively, the hardness of the computational task comes from the fact that calculating the summation on the right hand side of (\ref{previousres}) requires numerating $2^M$ combinations of $z$.

Now let us consider our ensemble case, specifically the circuit given in Fig. 3(a) of the main text.  For this calculation we shall derive the corresponding amplitude to (\ref{previousres}) in the qubit formalism.  Firstly the initial state can be written
\begin{align}
    | \psi_0 \rangle = \prod_{m=1}^M |0 \rangle^{\otimes N} = | 0 \rangle^{\otimes NM } .  
\end{align}
After the Hadamard gates the state becomes
\begin{align}
    H^{\otimes M } | \psi_0 \rangle = | + \rangle^{\otimes NM} = \frac{1}{\sqrt{2^{NM}}} \sum_{\vec{\sigma}} | \vec{\sigma} 
    \rangle , 
\end{align}
where 
\begin{align}
  \vec{\sigma} = (\sigma_{1,1}, \dots,  \sigma_{n,m}, \dots, \sigma_{N,M} ) 
\end{align}
is a microscopic spin configuration over the $ NM $ qubits.  The spin $ \sigma_{n,m} = \pm 1 $ refers to the spin configuration of the $ n$th spin within the $ m$th ensemble. Applying the sequence of gates in  Fig. 3(a) of the main text, we have

\begin{widetext}
\begin{align}
& \left[ \prod_{m} U_{m}^z (\theta_{m}) \right]
\left[ \prod_{m_1 m_2} T_{m_1 m_2} (\xi_{m_1 m_2}) \right]
\left[ \prod_{m_1 m_2 m_3} R_{m_1 m_2 m_3} (\chi_{m_1 m_2 m_3}) \right]   H^{\otimes M } | \psi_0 \rangle  \nonumber \\
& =\exp \left( -i \sum_{m}  \theta_{m}   \hat{S}^z_{m} \right)
\exp \left( -i \sum_{m_1 m_2}  \xi_{m_1 m_2}   \hat{S}^z_{m_1} \hat{S}^z_{m_2} \right)  \exp \left( -i \sum_{m_1 m_2 m_3}  \chi_{m_1 m_2 m_3}   \hat{S}^z_{m_1} \hat{S}^z_{m_2} \hat{S}^z_{m_3} \right) | + \rangle^{\otimes NM}  \nonumber \\
& = \frac{1}{\sqrt{2^{NM}}} \sum_{\vec{\sigma}} 
\exp \bigg(-i \sum_{m_1 m_2 m_3}  \sum_{n_1 n_2 n_3} \chi_{m_1 m_2 m_3}   \sigma_{n_1,m_1}  \sigma_{n_2,m_2}  \sigma_{n_3,m_3} 
-i \sum_{m_1 m_2} \sum_{n_1 n_2} \xi_{m_1 m_2}  \sigma_{n_1,m_1}  \sigma_{n_2,m_2} \nonumber \\
& -i \sum_{m} \sum_n \theta_{m}  \sigma_{n,m} \bigg) | \vec{\sigma}  \rangle  . 
\end{align}
\end{widetext}
The particular measurement outcome $ | 0 \rangle^{\otimes NM} $ has the amplitude
\begin{widetext}
\begin{align}
&  \langle 0 |^{\otimes NM}  H^{\otimes M } \left[ \prod_{m} U_{m}^z (\theta_{m}) \right]
\left[ \prod_{m_1 m_2} T_{m_1 m_2} (\xi_{m_1 m_2}) \right]
\left[ \prod_{m_1 m_2 m_3} R_{m_1 m_2 m_3} (\chi_{m_1 m_2 m_3}) \right]   H^{\otimes M } | \psi_0 \rangle   \nonumber \\
 & = \frac{1}{2^{NM}} \sum_{\vec{\sigma}} 
\exp \left( -i \sum_{m_1 m_2}  \xi_{m_1 m_2}   \hat{S}^z_{m_1} \hat{S}^z_{m_2} \right)  \exp \left( -i \sum_{m_1 m_2 m_3}  \chi_{m_1 m_2 m_3}   \hat{S}^z_{m_1} \hat{S}^z_{m_2} \hat{S}^z_{m_3} \right) | + \rangle^{\otimes NM}  \nonumber \\
& = \frac{1}{2^{NM}} \sum_{\vec{\sigma}} 
\exp \bigg( -i \sum_{m_1 m_2 m_3}  \sum_{n_1 n_2 n_3} \chi_{m_1 m_2 m_3}   \sigma_{n_1,m_1}  \sigma_{n_2,m_2}  \sigma_{n_3,m_3} 
-i \sum_{m_1 m_2} \sum_{n_1 n_2} \xi_{m_1 m_2}  \sigma_{n_1,m_1}  \sigma_{n_2,m_2} \nonumber \\
& -i \sum_{m_1} \sum_{n_1} \theta_{m_1}  \sigma_{n_1,m_1} 
\bigg) .  \label{amplitudefinal}
\end{align}
\end{widetext}
Let us now find what the coefficients $ \chi, \xi, \theta$ should be for a specified function $ f(\vec{\sigma}) $ given in Eq. (8) of the main text.  Substituting the relation 
\begin{align}
k_m = \frac{1}{2} (N + \sum_{n=1}^N \sigma_{n,m}),
\end{align}
into Eq. (8), we have
\begin{align}
f(\vec{\sigma})  = &
\sum_{m_1 m_2 m_3} \sum_{n_1 n_2 n_3} \frac{\alpha_{m_1 m_2 m_3}}{8}  \sigma_{n_1,m_1}  \sigma_{n_2,m_2}  \sigma_{n_3,m_3} \nonumber \\
& + \sum_{m_1 m_2} \sum_{n_1 n_2} \bigg[
\frac{N}{8} \sum_m ( \alpha_{m_1 m_2 m} + \alpha_{m_1 m m_2} \nonumber \\
& + \alpha_{m m_2 m_1} )+ \frac{\beta_{m_1 m_2}}{4} ) \bigg]  \sigma_{n_1,m_1}  \sigma_{n_2,m_2}  \nonumber \\
& + \sum_{m_1} \sum_{n_1}  \bigg[ \frac{N^2}{8} \sum_{m m'} ( \alpha_{m_1 m m'} + \alpha_{m m_1 m'} \nonumber \\
&+  \alpha_{m m' m_1}) + \frac{N}{4} \sum_m ( \beta_{m_1 m} + \beta_{m m_1}) + \frac{\gamma_{m_1}}{2} \bigg]  \sigma_{n_1,m_1} \nonumber \\
& + \frac{N^3}{8} \sum_{m_1 m_2 m_3} \alpha_{m_1 m_2 m_3}+ \frac{N^2}{4} \sum_{m_1 m_2} \beta_{m_1 m_2} \nonumber\\
&+ \frac{N}{2} \sum_{m_1} \gamma_{m_1} .  
\end{align}
Matching this to (\ref{amplitudefinal}), we may choose the circuit parameters as
\begin{align}
\chi_{m_1 m_2 m_3}  = & \frac{\pi \alpha_{m_1 m_2 m_3}}{8}  \nonumber \\
 \xi_{m_1 m_2}  = & \frac{N\pi }{8} \sum_m ( \alpha_{m_1 m_2 m} + \alpha_{m_1 m m_2} + \alpha_{m m_2 m_1} ) \nonumber \\
 &+ \frac{\pi \beta_{m_1 m_2}}{4}  \nonumber \\
\theta_{m_1} = & \frac{N^2 \pi }{8} \sum_{m m'} ( \alpha_{m_1 m m'} + \alpha_{m m_1 m'} +  \alpha_{m m' m_1}) \nonumber \\
&+ \frac{N\pi}{4} \sum_m ( \beta_{m_1 m} + \beta_{m m_1}) + \frac{\gamma_{m_1}\pi }{2} . 
\end{align}
Choosing these parameters, and taking the modulus squared of the amplitude (\ref{amplitudefinal}) gives the probability
\begin{align}
p_{\vec{0}}  = & |\langle 0 |^{\otimes NM}  H^{\otimes M } \left[ \prod_{m} U_{m}^z (\theta_{m}) \right]
\left[ \prod_{m_1 m_2} T_{m_1 m_2} (\xi_{m_1 m_2}) \right] \nonumber \\
&\times \left[ \prod_{m_1 m_2 m_3} R_{m_1 m_2 m_3} (\chi_{m_1 m_2 m_3}) \right]   H^{\otimes M } | \psi_0 \rangle|^2  \nonumber \\
= & \frac{1}{4^{NM}} | \sum_{\vec{\sigma}} e^{-i \pi f(\vec{\sigma})} |^2
\end{align}
as claimed in the main text.

We note that the sum over $ 2^{MN}$ can be reduced to a sum over $ (N+1)^M $ using the fact that $ k_m $ only depends on a global property of the ensemble.  This does not affect the complexity since even in the case of $ N = 1$, the problem is still \#P-hard. In the main text we have limited ourselves to $ \alpha, \beta, \gamma \in \{0,1\} $ in order to take advantage of the complexity of degree 3 polynomials.  Expanding the range of $\alpha,\beta,\gamma$ will give a larger range of values of $ e^{-i\pi f(\vec{\sigma}) } $, such that (7) in the main text will involve a sum over $(N+1)^M$ complex phases without any structure, further increasing the complexity of the evaluation. 

\subsection{ Worst-to-average-case reduction }
In this part we provide a sketch of proof for the worst-to-average-case reduction developed in Ref. \cite{Bouland2018} which enables us to prove \#P-hardness of exactly computing the output probability of an average-case quantum circuit. In the previous part we have shown the existence of worst-case circuit whose output probability is \#P-hard to exactly compute. By showing that the probabilities of worst-case circuit can be obtained from probabilities of average-case circuit in polynomial time, we can prove that the exact computation of the latter is also \#P-hard. The basic idea is to express the probability as a function of an variable characterizing different circuits. When the variable takes certain values, the function outputs the probability of the worst-case circuit. If we can infer the form of the function using polynomial number of points obtained from the average-case circuits, the worst-to-average reduction is established.

The first step is to to connect the average-case circuit and the worst-case circuit based on the Haar-measure invariance of matrix multiplication. Specifically, we take a worst-case circuit $\mathcal{C}$ and a haar-random matrix $\mathcal{H}$, and construct a circuit as the multiplication of $\mathcal{C}$ and a fraction of $\mathcal{H}$: 
\begin{align}
     &\mathcal{G}_1(\theta) = \mathcal{C} \mathcal{H}_1(\theta) \label{C1}
     \\
     &\mathcal{H}_1(\theta) = \mathcal{H} \mathrm{e}^{-\theta\log \mathcal{H}}
\end{align}
When $\theta=1$, $\mathcal{G}_1(1) = \mathcal{C}$ which becomes the worst-case circuit, but if $\theta\rightarrow0$, the circuit is completely scrambled, in this way $\mathcal{G}_1(\theta)$ builds the connection between worst case and average case via the variable $\theta$.

Now the task becomes using $\langle\vec{0}|\mathcal{G}_1(\theta\rightarrow 0)|\psi_0\rangle $ to calculate $\langle\vec{0}|\mathcal{G}_1(\theta=1)|\psi_0\rangle $ in polynomial time. Through the Berlekamp-Welch Algorithm\cite{welch1986} we know that if $\langle\vec{0}|\mathcal{G}_1(\theta)|\psi_0\rangle$ is a degree $d$ polynimial of $\theta$, then with a least $d+1$ different points, $\langle\vec{0}|\mathcal{G}_1(\theta)|\psi_0\rangle$ can be recovered in poly($d$) deterministic time. Therefore the next step is to Taylor expand $\mathrm{e}^{-\theta\log \mathcal{H}}$ in Eq.(\ref{C1}) and truncate at certain degree: 
\begin{align}
    &\mathcal{G}_2(\theta,K) = \mathcal{C} \mathcal{H}_2(\theta,K)  \label{C2}
    \\
    &\mathcal{H}_2(\theta,K) = \mathcal{H} \sum_{k=0}^K \dfrac{(-\theta \log \mathcal{H})^k}{k!}
\end{align}
According to the standard bound of Taylor series, the distance between $\langle\vec{0}|\mathcal{G}_2(\theta,K)|\psi_0\rangle$ and $\langle\vec{0}|\mathcal{G}_1(\theta)|\psi_0\rangle$ is at most $2^{-\text{poly}(n)}$ for a sufficiently large choice of $K=\text{poly}(n)$. This applies for the distance between $\langle\vec{0}|\mathcal{G}_2(1,K)|\psi_0\rangle$ and $\langle\vec{0}|\mathcal{C}|\psi_0\rangle$ as well which is what we wish to calculate in the end.

As a Haar-random matrix, $\mathcal{H}$ can be decomposed into its diagonal form as
\begin{equation}
    \mathcal{H} = U \text{diag}( \mathrm{e}^{\mathrm{i}\phi_1}, \mathrm{e}^{\mathrm{i}\phi_2}, ..., \mathrm{e}^{\mathrm{i}\phi_N} ) U^\dag
\end{equation}
where $U$ is formed with eigenvectors of the decomposition. Therefore $\mathrm{e}^{-\theta\log \mathcal{H}}$ can only affect the eigenvalue of $\mathcal{H}$, reducing the range of $\{\phi_1, \phi_2,..., \phi_N\}$ from $2\pi$ to $2\pi(1-\theta)$. For a Haar-random choice of $\mathcal{H}_1$, with a probability $1 - 1/\text{poly} (n) $ it falls into the distribution of constructed average-case circuit $\mathcal{G}_1(\theta)$ if we let $\theta$ = 1/poly($n$).

Combining these two steps it is easy to see that if there exists a machine \textit{O} that can exactly compute $\langle \vec{0} |\mathcal{G}_2|\psi_0\rangle$ in polynomial time for 3/4 of $G_2$ from the aforementioned distribution as an average-case circuit, then with at least $d+1$ queries of $O$ for $\{\theta_1, \theta_2,..., \theta_{d+1}\}\in[0, 1/\text{poly}(n))$, the degree $d$ polynomial $\langle\vec{0}|\mathcal{G}_2(\theta,K) |\psi_0\rangle$ can be recovered using the Berlekamp-Welch Algorithm. After that we can calculate  $\langle \vec{0}|\mathcal{G}_2(1,K) |\psi_0\rangle$ which equals to $\langle \vec{0}|\mathcal{C}|\psi_0\rangle$ within additive precision $2^{-\text{poly}(n)}$. In this way we have reduced the hardness of exactly computing the output probability of the worst-case circuit to that of the average-case circuit.

\subsection{The complexity of approximating the average-case circuit}

In previous part we have sketched the worst-to-average-case reduction for exact computation. In this part, we show that this reduction still works for the approximation version of the problem. This is mainly because the distance between the worst-case probability $\langle\vec{0}|\mathcal{C}|\psi_0\rangle$ and $\langle\vec{0}|\mathcal{G}_2(1,K)|\psi_0\rangle$ which is calculated through average-case probabilities is much smaller than the distance between $\langle\vec{0}|\mathcal{C}|\psi_0\rangle$ and its approximation $\widetilde{\langle\vec{0}|\mathcal{C}|\psi_0\rangle}$ that $\left| \langle\vec{0}|\mathcal{C}|\psi_0\rangle-\langle\vec{0}|\mathcal{G}_2(1,K)|\psi_0\rangle \right| < \left|\langle\vec{0}|\mathcal{C}|\psi_0\rangle-\widetilde{\langle\vec{0}|\mathcal{C}|\psi_0\rangle}\right| \leq 2^{-n}/\text{poly}(n) $. Hence if there exists an efficient approximation method for average-case circuit, using the same reduction technique in previous part the approximated probability for the worst-case circuit can also be calculated. 

In other words, if there exists a $(\delta,\epsilon)$-approximator $O'$ for the average-case circuit:
\begin{equation}
    \text{Pr}\left( | O(\mathcal{G}_2(\theta,K)) - \langle\vec{0}|\mathcal{G}_2(\theta,K)|\psi_0\rangle | \leq \epsilon \right) \geq 1 - \delta
\end{equation}
It is then a $(\delta',\epsilon')$-approximator $O'$ for the worst-case circuit:
\begin{equation}
    \text{Pr}\left( | O(\mathcal{G}_2(1,K)) - \langle\vec{0}|\mathcal{C}|\psi_0\rangle | \leq \epsilon' \right) \geq 1 - \delta'
\end{equation}
where $\delta'=\delta+1/\text{poly}(n), \epsilon'=\epsilon+1/\text{exp}(n)$.

\subsection{From approximating individual output probabilities to classical simulation}

So far we have established the complexity of approximating the single output probability of an average-case circuit. Now we move on to disprove the existence of efficient classical sampler that can simulate the distribution to variation distance errors.

Let us suppose there exists an efficient classical sampler $\mathcal{A}(\mathcal{G}')$ that can sample from a probability distribution which approximates the output probability distribution of an average-case circuit $\mathcal{G}'$ up to additive $\epsilon$ in $l_1$ norm
\begin{equation}
    \sum_{\vec{x}} |q_{\vec{x}} - p_{\vec{x}}| < \epsilon,
\end{equation}
where $q_{\vec{x}}$ denotes the probability of obtaining output $\vec{x}$ from $\mathcal{A}(\mathcal{G}')$. According to the Stockmeyer Counting Theorem, there exists an algorithm with access to an NP-oracle that can approximate $q_{ \vec{x} }$ to a multiplicative error:
\begin{equation}
    |\widetilde{q}_{\vec{x}} - q_{\vec{x}} | \leq \dfrac{q_{\vec{x}}}{\text{poly}(n)}.
\end{equation}

Since $p_{\vec{x}}$ follows the Porter-Thomas distribution such that
$\text{E}_{\vec{x}}(p_{\vec{x}}) = 1/(N+1)^M$, then from Markov's inequality we have
\begin{equation}
    \text{Pr}\left( |q_{\vec{x}}-p_{\vec{x}} | \geq \dfrac{\epsilon}{(N+1)^M\delta}  \right) \leq \delta.
\end{equation}
Then with probability at least $1-\delta$ over the choice of $\vec{x}$,
\begin{equation}
    \begin{split}
        |\widetilde{q}_{\vec{x}} - p_{\vec{x}} | &\leq |\widetilde{q}_{\vec{x}} - q_{\vec{x}} | + |q_{\vec{x}}-p_{\vec{x}} |
        \\
        &\leq \dfrac{q_{\vec{x}}}{\text{poly}(N)} + |q_{\vec{x}}-p_{\vec{x}} |
        \\
        &\leq \dfrac{p_{\vec{x}}}{\text{poly}(N)} + \left(1+\dfrac{1}{\te{poly}(N)}\right) |q_{\vec{x}}-p_{\vec{x}} |
        \\
        &\leq \dfrac{p_{\vec{x}}}{\text{poly}(N)} + \left(1+\dfrac{1}{\te{poly}(N)}\right)\dfrac{\epsilon}{(N+1)^M\delta} .  
    \end{split}
\end{equation}
Letting $\epsilon=\delta/8$ and Porter-Thomas distribution gives 
\begin{equation}
    \te{Pr}\left( p_{\vec{x}} > \dfrac{1}{(N+1)^M} \right) = \dfrac{1}{e},
\end{equation}
then with probability $1/e - \delta$, we can approximate $p_{\vec{x}}$ to a multiplicative error $1/4 + o(N)$.

Therefore the existence of an efficient classical sampler indicates the existence of a $\te{BPP}^\te{NP}$ algorithm that can approximate the output probability of average-case circuit which is proven to be a \#P-hard problem. This will collapse the polynomial hierarchy to the third order because $\te{BPP}^\te{NP}$ is on the third order of the polynomial hierarchy \cite{arora_computational_2009} yet the whole polynomial hierarchy is contained in \#P-hard \cite{toda_pp_1991}.

\section{Feynman Path Integral based sampling algorithm}

\subsection{Random qubit circuits}

We first review the methods shown in Refs. \cite{Boixo2017,Boixo2018}, which showed the Feynman Path Integral (FPI) based classical sampling algorithm for simulating a qubit-based quantum circuit. 

According to the definition of path integral, the amplitude of finding the output of a particular qubit circuit can be expressed as
\begin{equation}
    \langle \vec{\sigma}_f|{\hat{\cal C}} | \psi_0\rangle = \sum_{  \vec{\sigma}^1, \dots , \vec{\sigma}^{T-1} } \prod_{t=1}^T \langle \vec{\sigma}^t | C^{(t)} | \vec{\sigma}^{t-1}\rangle, 
\end{equation}
where $ | \vec{\sigma}^t \rangle = \otimes_{m=1}^M |\sigma_m^t\rangle$ is a spin configuration and $\sigma_m^t = \pm 1$. The initial state is $ | \vec{\sigma}^{0} \rangle =  | \psi_0\rangle $.  The number of the gates applied in total is $T$ and $| \vec{\sigma}_f \rangle = |\vec{\sigma}^T \rangle$ represents the final state. The total circuit consists of the sequence
\begin{align}
{\hat{\cal C}} = \prod_{t=1}^T  C^{(t)} , 
\label{gatebygate}
\end{align}
where the product is applied in order from right to left, from $ t = 1 $ to $ t = T$. In (\ref{gatebygate}), each of the $ C^{(t)} $ is a single gate, such that the total circuit written gate by gate.

\begin{figure}[t]
\includegraphics[width=8cm]{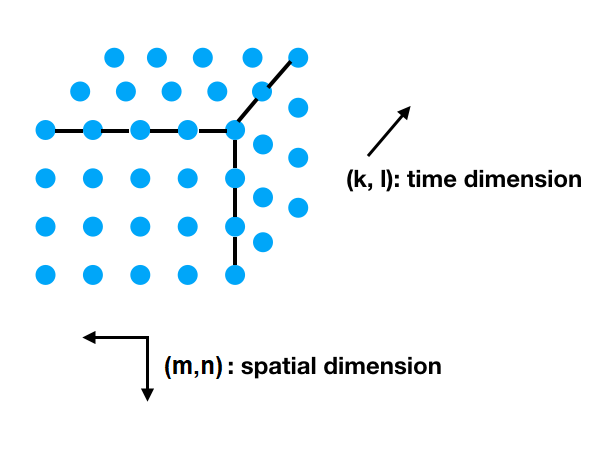}
\caption{Mapping of the quantum circuit to a classical Ising model.  }
\label{Ising}
\end{figure}

Ref. \cite{Boixo2017} shows that it is equivalent to evaluating
\begin{equation}
   \langle \vec{\sigma}_f|{\hat{\cal C}} | \psi_0\rangle = 2^{-G/2}\sum_s \text{exp}(i  H_s ) ,
   \label{amplitudepathint}
\end{equation}
where $G = \sum_{m=1}^M d(m)$ is the total number of two-sparse gates, i.e. gates with non-zero off diagonal terms. In a given path qubit $m$ goes through the path $\{s_m^k\}_{k=0}^{d(m)}$, where $d(m)$ is the number of two-sparse gates applied to qubit $m$.  Then the set of all possible path is  $s = \{s_m^k\}$ with $m\in[1\ldots M]$ and $k\in[0\ldots d(m)-1]$. The total number of paths required to recover the exact probability distribution is $2^G$. Here, $H_s$ is a Hamiltonian describing an effective classical 3D Ising model for the path integral model, which is given by
\begin{equation}
H_s = \sum_{m=1}^M \sum_{k=1}^{d(m)-1}h_m s_m  +\sum_{m'<m}^M  \sum_{k=1}^{d(m)-1}\sum_{l=1}^{d(m')-1} \mathcal{J}_{m m'}^{kl}s_m^k s_{m'}^l
\label{fullIsing}
\end{equation}
where $m,m'$ denotes the space degree of freedom that decides the coupling between different spins and $k,l$ denotes the time degree of freedom that constrains the layers of interaction as shown in Fig. \ref{Ising}. We note that $ H_s $ also depends upon upon the initial $ | \psi_0 \rangle $ and final state $ | \vec{\sigma}_f \rangle $ through boundary terms which are of fixed spin configuration $ s_m^k$.

\subsection{Multi-ensemble random quantum circuit}

We now apply the same technique to derive the path integral calculation for our ensemble based random quantum circuit.  As before, we would like to calculate the amplitude
\begin{align}
\langle \vec{k}_f | \hat{{\cal C}} | \psi_0 \rangle = 
\sum_{ \vec{k}^1, \dots, \vec{k}^{T-1} } \prod_{t=1}^T 
\langle \vec{k}^{t} | C^{(t)}| \vec{k}^{t-1} \rangle , 
\label{pathintens2}
\end{align}
where $ | \vec{k}^t \rangle= \otimes_{m=1}^M | k_m^t \rangle $ is a ensemble configuration with $ k_m^t \in [0,N] $.  The initial state is $ | \vec{k}^0 \rangle = | \psi_0 \rangle $ and the final state is $ | \vec{k}_f \rangle = | \vec{k}^T \rangle $.   The total circuit consists of the sequence
\begin{align}
{\hat{\cal C}} =  \left( \prod_{l=1}^L \left[ \prod_{m=1}^M W_m^{(l)} \right]  \hat{\cal Q} \right)  \left( \prod_{m'=1}^M H_{m'} \right) , 
\end{align}
where we denote the basis rotation operators $ W_m^{(l)} $ as the random rotation from the choice $\{ \hat{X}_m^{1/2}, \hat{Y}_m^{1/2}, \hat{Z}_m^{1/4} \} $ on the $ l$th cycle.  
The product is applied in order from right to left, from $ l = 1 $ to $ l = L$. The remaining products  labels are also taken to be in order from right to left as $ m , m' $ increases, although the terms within these product commute so this is an arbitrary choice.  The $ C^{(t)} $ operators are then the individual gates in this order of this sequence, for example
\begin{align}
C^{(1)} & = H_1 \nonumber \\
C^{(M)} & = H_M \nonumber \\
C^{(M+1)} & = {\cal Q}  \nonumber \\
C^{(M+2)} & = W_1^{(1)}  \nonumber \\
C^{(2M+1)} & = W_M^{(1)}  \nonumber \\
C^{(2M+2)} & = {\cal Q} .  
\end{align}

The matrix elements in (\ref{pathintens}) are diagonal for the operators $ \cal Q $ and $ Z_m^{1/4} $ taking matrix elements
\begin{align}
  \langle \vec{k} | {\cal Q} |  \vec{k'} \rangle & = \delta_{\vec{k} \vec{k'} } e^{-i \sum_{m=1}^M (2k_m- N)^2 \xi } \nonumber \\ 
\langle \vec{k} | Z_m^{1/4} |  \vec{k'} \rangle & = \delta_{\vec{k} \vec{k'} } e^{-i (2k_m-N) \pi/8 } .  
\label{diagmatelem}
\end{align}
The two remaining off-diagonal (two-sparse) operators have matrix elements
\begin{align}
 \langle \vec{k} | X_m^{1/2} |  \vec{k'} \rangle & =
\left( \prod_{m'\ne m } \delta_{k_{m'} k_{m'}'}  \right) \langle k_m | e^{-i S_m^y \pi/4 } | k_m' \rangle e^{i(k_m' -k_m)\pi/2} \nonumber \\
 \langle \vec{k} | Y_m^{1/2} |  \vec{k'} \rangle & =
\left( \prod_{m'\ne m } \delta_{k_{m'} k_{m'}'}  \right) \langle k_m | e^{-i S_m^y \pi/4 } | k_m' \rangle , \label{offdiagmatelem}
\end{align}
where \cite{Byrnes2021}
\begin{align}
& \langle k | e^{-i S^y \pi/4} | k' \rangle = \frac{\sqrt{ k'! (N-k')! k! (N-k)!} }{\sqrt{2^N}} \nonumber \\
& \times \sum_n \frac{(-1)^n  }{(k-n)!(N-k'-n)!n!(k'-k+n)!} .   
\label{syrotmatrixelement}
\end{align}

We may picture the multidimensional sum in (\ref{pathintens}) as a path through configurational space, where there are $ (N+1)^{M(T-1)} $ different routes to get from the initial state $ |\vec{k}^0 \rangle $ to $ |\vec{k}^{T} \rangle $.  Due to the  delta functions in the matrix elements (\ref{diagmatelem}) and (\ref{offdiagmatelem}), many of these paths have zero amplitude and hence may be removed.  Only the off-diagonal terms (\ref{offdiagmatelem}) create a branching in configurational space, hence the total number of paths reduces to $ (N+1)^G $, where $ G $ is the total number of two-sparse ensemble gates.  Running over the complete set of these paths in (\ref{pathintens}) gives the final result.  

Summing over all $ (N+1)^G $ paths may be exceedingly large to perform exhaustively. In this case, we may approximate the path integral by a random sample of the paths.  Starting from the configuration corresponding to the initial state $ \vec{k}^0 $, we choose a random new configuration each time one of the two-sparse gates are applied.  For example, for  a two-sparse gate on ensemble $ m $, the new configuration is chosen according to
\begin{align}
\vec{k} = (k_1,\dots,  k_m, \dots, k_M)
\rightarrow
\vec{k}' = (k_1,\dots,  k_m', \dots, k_M)
\end{align}
where $ k_m' $ is randomly selected from a uniform distribution.
For the diagonal matrices $ Z_m^{1/4} $ and $ \cal Q$, the configuration is unchanged. This is repeated until the $ (T-1)$th configuration.  The final configuration is fixed to $ | \vec{k}_f \rangle $, according to the final state.  We then estimate the amplitude according to 
\begin{align}
\Psi_{\text{est}} (\vec{k}_f) = \langle \vec{k}_f | \hat{{\cal C}} | \psi_0 \rangle_{\text{est}} \propto
\sum_{ \vec{k} \in {\cal K} } \prod_{t=1}^T 
\langle \vec{k}^{t} | C^{(t)}| \vec{k}^{t-1} \rangle , 
\label{pathintest}
\end{align}
where $ {\cal K} $ is a set of random paths in configuration space. 

To find the fidelity, we perform the procedure in (\ref{pathintest}) for all the output states $ |  \vec{k}_f \rangle $.  We then normalize the state and compare it to the exact value using the fidelity
\begin{align}
F = \left| \sum_{\vec{k} } \Psi_{\text{est}}^* (\vec{k}_f) \Psi_{\text{exact}} (\vec{k}_f) \right|^2 .  
\end{align}
The exact wavefunction is evaluated by a direct matrix multiplication 
\begin{align}
\Psi_{\text{exact}} (\vec{k} ) =  \langle \vec{k} |    {\cal C} | \psi_0 \rangle . 
\end{align}

\end{document}